\documentclass[pra,aps,twocolumn,tightenlines,showpacs,amssymb,amsfonts,superscriptaddress]{revtex4}

\usepackage{epsfig}

\begin{document}

\title{Multipartite entanglement and quantum state exchange}
\author{D. T. Pope}
\affiliation{School of Physical Sciences \& Centre for Quantum
Computer Technology, University of Queensland, Brisbane 4072,
Queensland, Australia} \affiliation{Centre for Quantum Dynamics,
School of Science, Griffith University, Nathan 4111, Queensland,
Australia} \email{d.pope@griffith.edu.au}
\author{G. J. Milburn}
\affiliation{School of Physical Sciences \& Centre for Quantum
Computer Technology, University of Queensland, Brisbane 4072,
Queensland, Australia} \email{milburn@physics.uq.edu.au}
\date{ }

\begin{abstract}
We investigate multipartite entanglement in relation to the
process of quantum state exchange. In particular, we consider such
entanglement for a certain pure state involving two groups of $N$
trapped atoms. The state, which can be produced via quantum state
exchange, is analogous to the steady-state intracavity state of
the subthreshold optical nondegenerate parametric amplifier. We
show that, firstly, it possesses {\em some} $2N$-way entanglement.
Secondly, we place a lower bound on the amount of such
entanglement in the state using a novel measure called the {\em
entanglement of minimum bipartite entropy}.
\end{abstract}


\maketitle

\section{Introduction}

%
%what is multipartite entanglement?
%

Multipartite entanglement is entanglement that crucially involves
three or more particles. A well-known example of it occurs in the
generalized GHZ state $| \psi \rangle = |0  \rangle^{\otimes M} +
|1 \rangle^{\otimes M}$, where $M$ is an integer greater than two.
Multipartite entanglement is an interesting quantum resource for a
number of reasons. Firstly, it is a key resource in quantum
computation as i) it has been proved that it is a necessary
ingredient in order for a quantum computation to obtain an
exponential speed-up over classical computation \cite{jozsa01} and
ii) it is central to quantum error correction which uses it to
encode states, to detect errors and, ultimately, to help implement
fault-tolerant quantum computation (see, for example, Chapter 10
of \cite{nielsen00}). The second reason why multipartite
entanglement is interesting is that it can manifest nonclassical
correlations such as GHZ-type correlations \cite{zukowski99}.
Finally, it has been conjectured that multipartite entangled
states contain a wealth of interesting and unexplored physics
\cite{preskill}.

%
%previous multipartite entanglement measures
%

In order to quantify the amount of multipartite entanglement
present in a state, a number of measures have been proposed.
Firstly, Vedral {\em et al.} \cite{vedral97} have suggested a
measure of multipartite entanglement for a state $\rho$ which is
the minimum relative entropy between $\rho$ and any separable
state. For systems with more than two subsystems, they defined a
separable state as one in which the state of at least one
subsystem can be factored out from that of the others. In
addition, Coffman, Kundu and Wootters \cite{coffman00} have
extended the bipartite entanglement measure called the tangle
\cite{wootters98} to the {\em 3-tangle} which measures 3-way
GHZ-type entanglement. Furthermore, Vidal \cite{vidal00} has
studied {\em entanglement monotones} --- quantities whose
magnitudes do not increase on average under local transformations
--- and has proposed that {\em all} of them can be regarded as
entanglement measures. Meyer and Wallach \cite{meyer01} have
proposed a measure of ``global entanglement'' for $n$-qubit pure
states which is the sum of a number of terms involving wedge
products. Each wedge product involves computational-basis
expansion coefficients for various $(n-1)$-qubit states obtained
by deleting a qubit from the state of interest. Similarly, Wong
and Anderson \cite{n_tangle} have extended the tangle to an
arbitrary even number of qubits for pure states. Finally, Biham,
Nielsen and Osborne \cite{biham01} have proposed the {\em
Groverian entanglement} for a pure state $| \psi \rangle$ based on
how successful Grover's algorithm \cite{grover96} performs given
the input $| \psi \rangle$. The Groverian entanglement is
equivalent to a measure in \cite{vedral97}, however, it shows an
interesting link between an entanglement measure and the quantum
information processing capability of states. In addition to the
measures listed above a number of others have also been proposed.

%
%what is quantum state exchange?
%

{\em Quantum state exchange}
\cite{parkins1,parkins2,parkins4,parkins3,parkins5} is a newly
formulated process by which information is transferred from an
electromagnetic field to the vibrational state of one or more
trapped atom(s). It is implemented using a stimulated Raman
process that couples the electromagnetic field to the vibrational
state and thus transfers information from the former to the
latter. We explain it in more detail in
Section~\ref{background_theory}.

%
%paper summary
%

In this paper we show that quantum state exchange can be used to
create an entangled state for $2N$ trapped atoms that is a useful
quantum resource. We begin in,
Subsection~\ref{quantum_state_exchange}, by explaining the process
of quantum state exchange via presenting a detailed example of it
within a simple system consisting of a harmonically trapped atom
interacting with a cavity mode. Next,
Subsection~\ref{state_of_interest} shows that quantum state
exchange can be used to generate an entangled pure state for two
groups of $N$ trapped atoms located in two spatially separated
far-off-resonance dipole-force traps (FORTs). In
Subsection~\ref{summary} we present a detailed summary of the
remainder of the paper. Sections~\ref{section_qual} and
\ref{ch5:quant} then investigate the nature of the state's
$2N$-way entanglement. That is, the nature of its entanglement
that spans across all $2N$ atoms. In particular, in
Subsections~\ref{npt} and \ref{ch5:quant} we {\em qualitatively}
explore this entanglement by presenting a novel necessary and
sufficient condition for the presence of $M$-way entanglement for
$M$-partite pure states and then showing that the state satisfies
it. In Section~\ref{ch5:quant} we {\em quantitatively} explore the
state's $2N$-way entanglement by first presenting a novel
multipartite entanglement measure for pure states in
Subsection~\ref{quant_theory}. This measure is based on the von
Neumann entropies \cite{nielsen00,preskill98} for all the reduced
density operators obtainable from some pure state of interest by
tracing over some of the subsystems for the state. After defining
the measure, we then use it to calculate lower bounds on the
amount of $2N$-way entanglement in the state in
Subsection~\ref{quant_results}. Finally, we discuss our results in
Section~\ref{discuss}.

\section{Background theory} \label{background_theory}

\subsection{Quantum state exchange} \label{quantum_state_exchange}

\begin{figure}[ht]
\center{\epsfig{figure=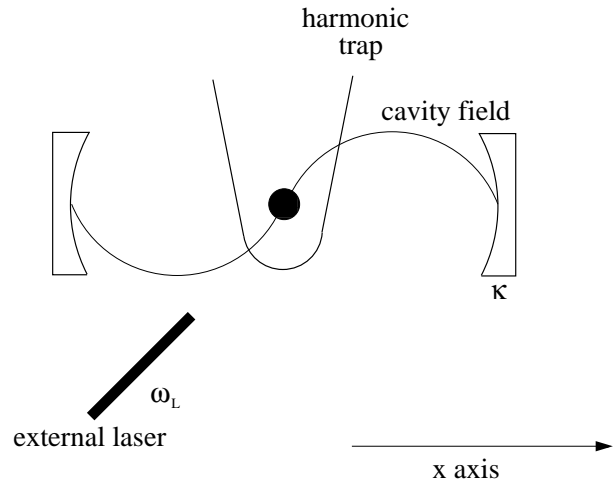,width=80mm}}
\caption{\label{ch3:fig1} Schematic diagram of a simple system in
which quantum state exchange can occur. A two-level atom
(represented by a black circle) lies within a harmonic trap that
is itself inside an optical cavity. The cavity, which is aligned
along the $x$-axis, supports a mode of frequency $\omega_{c}$ and
has one lossy mirror (with damping constant $\kappa$) and one
ideal one. An external laser of frequency $\omega_{L}$ is incident
from a direction perpendicular to the $x$-axis.}
\end{figure}

Perhaps the simplest system in which quantum state exchange can
occur \cite{parkins1} involves a two-level atom confined within a
harmonic trap which, in turn, lies inside a linearly damped
optical cavity with one lossy mirror and one ideal one. The atom's
vibrational motion is described by the annihilation operators
$b_{x},b_{y}$ and $b_{z}$. The two-level atom, which has a
transition frequency of $\omega_{a}$, couples to both an
intracavity electromagnetic field mode of frequency $\omega_{c}$
described by the annihilation operator $a$ and an external laser
of frequency $\omega_{L}$. The cavity and external laser
frequencies are chosen so as to drive Raman transitions that
couple adjacent atomic vibrational levels. Furthermore, the
cavity's axis coincides with the $x$-axis whilst the external
laser's beam is perpendicular to this axis. Lastly, we assume that
the harmonic trap is centred on a cavity-field node and thus a
schematic diagram of this system is as in Fig.~\ref{ch3:fig1}.

%
%hamiltonian
%

The system's Hamiltonian is, in a reference frame
rotating at frequency $\omega_{L}$,
\begin{eqnarray} \label{hamiltonian}
H_{\rm total}^{\rm single} & = &   H_{\rm sys} + \kappa (a
R^{\dag} + a^{\dag} R) +  H_{\rm res},
\end{eqnarray}
where $H_{\rm res}$ is the free Hamiltonian of the reservoir coupled to
the cavity mode, $\kappa$ is a damping rate, $R$ is a reservoir operator and
$H_{\rm sys}$ is the Hamiltonian for the cavity-atom system which is
\begin{eqnarray} \label{single_atom} \nonumber
H_{\rm sys} & = &
\Sigma_{j=x,y,z} \hbar \nu_{j} (b^{\dag}_{j} b_{j} + 1/2)
+ \hbar \delta a^{\dag} a + \hbar \Delta \sigma_{+} \sigma_{-} \\ \nonumber
&  & + \hbar \big[  {\cal E}_{L}(y,z,t) \sigma_{+} +  {\cal
E}^{*}_{L}(y,z,t) \sigma_{-} \bigr] \\
& & + \hbar g_{0} \sin (k x) \bigl( a^{\dag} \sigma_{-}   +   a \sigma_{+} \bigr),
\end{eqnarray}
where $\nu_{x},\nu_{y}$ and $\nu_{z}$ are the harmonic-oscillator frequencies
along the trap's $x$, $y$ and $z$ axes, $\sigma_{+}$ and $\sigma_{-}$
are atomic raising and lowering operators for the two-level atom,
$\delta=\omega_{c} -\omega_{L}$, $\Delta = \omega_{a} -\omega_{L}$,
${\cal E}_{L}$ is the complex
amplitude of the external laser field, $k= 2\pi/\lambda$, where $\lambda = 2 \pi c/\omega_{c}$,
$x = \sqrt{\hbar / 2 m \nu_{x} } ( b_{x}+ b_{x}^{\dag} ) $,
where $m$ is the mass of the two-level atom, and $g_{0}$ ($g_{0} \in \mathbb{R}$) is the coupling constant for the
atom-field interaction. Observe that $\omega_{c} -\omega_{L} = \nu_{x}$.

%
% simplifying assumptions made
%

The following reasonable assumptions are made about the system so as
to make calculations involving it more tractable:
\begin{enumerate}
\item The cavity field and external laser frequencies are
appreciably detuned from $\omega_{a}$ and
the two-level atom is initially in the ground state.
Thus, the excited internal state
is sparsely populated and spontaneous emission effects are negligible
and can be ignored.
\item Vibrational decoherence occurs over a timescale much longer than that of the
interaction producing quantum state exchange, as is the case
in an ion-trap realization of the system
\cite{parkins1}. Consequently, it has
a minimal effect over our timescale of interest and is ignored.
\item The trap dimensions are small compared to the cavity mode wavelength and thus
$\sin (k x) \ll 1$.
It follows from this that $\sin k x \simeq \eta_{x} (b_{x} + b_{x}^{\dag})$,
where $\eta_{x} = k \sqrt{\hbar/2m \nu_{x}}$. It also follows that
we can arrange things so that the $y$ and $z$ dependence of the external
laser field is negligible and thus, assuming ${\cal E}_{L}$ is time independent,
that ${\cal E}_{L}(y,z,t) \simeq {\cal E} e^{-i\phi_{L}}$, where
${\cal E}$ is a real time-independent amplitude.
\item The damping parameter $\kappa$ is such that $\nu_{x} \gg \kappa \gg g_{0}\eta_{x} {\cal E}/\Delta$.
\end{enumerate}

%
%how this system instansiates qusex
%

Given the assumptions above, and also assuming that
$g^{2}_{0}/\Delta \ll \delta$ and $g_{0}\langle a^{\dag} a\rangle
\ll {\cal E}_{L}$, $H_{\rm total}^{\rm single}$ can be rewritten
as \cite{parkins5} (p. 012308)
\begin{eqnarray} \label{adiabatic_elim} \nonumber
H_{\rm total}^{\rm single} &=  &  \Sigma_{j=x,y,z} \hbar \nu_{j}
\left( b_{j}^{\dag}b_{j} + 1/2 \right) + \hbar \delta a^{\dag} a -
\frac{\hbar {\cal E}^{2}}{\Delta} \\ \nonumber & & - \frac{\hbar
g_{0} {\cal E}}{\Delta} \sin(kx)\left(a^{\dag} e^{-i \phi_{L}} + a
e^{i
\phi_{L}}\right)  \\
& & + \kappa (a R^{\dag} + a^{\dag} R) + H_{\rm res}
\end{eqnarray}
by adiabatically eliminating the evolution of the internal states.
Furthermore, for the system under consideration, it has been shown
\cite{parkins1} that, in the steady-state regime, the vibrational
state of the atom in the $x$ direction is solely determined by the
input field (i.e. the light field entering the cavity) such that
\begin{equation} \label{indicate_qusex}
\tilde{b}_{x}(\omega) = \frac{\sqrt{2\Gamma}}{i\omega -\Gamma} \tilde{a}_{\rm in}(\omega),
\label{qusex_eqn}
\end{equation}
where $\tilde{b} (\omega) =
\frac{1}{\sqrt{2\pi}}\int_{t=-\infty}^{\infty} dt b_{x}(t) e^{ i \left(
\nu_{x} - \omega \right) t}$,
$\Gamma = g_{0}^{2}  \eta_{x}^{2} {\cal E}^{2} / \left( \Delta^{2} \kappa \right)$
and $\tilde{a}_{\rm in}(\omega) = \frac{1}{\sqrt{2 \pi}} \int_{t=-\infty}^{\infty} dt
e^{-i\omega t} \tilde{a}_{\rm in}(t)$,
where $\tilde{a}_{\rm in} (t) = \frac{ 1 }{ \sqrt{2\pi} }
\int_{ \omega'= -\infty }^{\infty } d\omega' e^{i \left(
\nu_{x}  - \omega' \right) t} c_{0} (\omega')$,
where $c_{0}(\omega')$ is the value of the reservoir annihilation
operator for the frequency $\omega'$ at time $t=0$.
The proportionality between $\tilde{b}_{x}(\omega)$ and $\tilde{a}_{\rm in}(\omega)$
present in Eqn~(\ref{indicate_qusex}) denotes that the
``statistics of the input field [have been] \ldots `written onto' the
state of the oscillator'' \cite{parkins1} (p. 498) (i.e. onto the atom's vibrational state in the
$x$ direction). We thus say that quantum state exchange has taken place when this equation holds.

\subsection{System of interest} \label{state_of_interest}

%
%\psi_{\rm CM}
%

In this paper we use quantum state exchange to generate a
particular state involving two groups of $N$ trapped atoms which
we later show contains multipartite entanglement that is a useful
quantum resource. Our work follows on from \cite{parkins2} in
which it was {\em asserted} that for a certain system consisting
of two groups of $N$ trapped atoms, quantum state exchange could
generate ``a highly entangled state of all 2[N] atoms". The system
we consider can be seen as a concrete example of that described in
the last paragraph of \cite{parkins2} --- our main original
contributions are, firstly, {\em demonstrating} that the state
within our system is a multipartite entangled one and, secondly,
quantitatively analysing the multipartite entanglement within it.

\begin{figure}[ht]
\center{\epsfig{figure=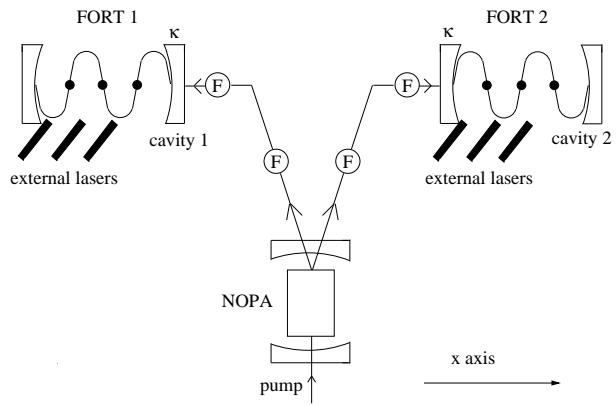,width=80mm}}
\caption{\label{ch3:fig_two} Schematic diagram for a system
involving, firstly, a subthreshold optical nondegenerate
parametric amplifier (NOPA) whose output modes pass through
Faraday isolators (represented by an F enclosed in a circle) and
then feed into linearly damped optical cavities (as indicated by
the sinusoidal curves inside both cavites which represent the
cavity modes $a_{1}$ and $a_{2}$). These cavities are aligned
along the $x$-axis and both have one ideal mirror and one lossy
one (with damping constant $\kappa$). Inside each of them is a
far-off-resonance dipole-force trap (FORT) that, along with a
two-dimensional far-off resonance optical lattice, confines $N$
identical two-level atoms in linear chain parallel to the
$x$-axis. Observe that the FORTs' trapping modes are not shown in
the diagram. External lasers of frequency $\omega_{L}$ are
incident on each atom in both traps from a direction perpendicular
to the $x$-axis. }
\end{figure}

The system we consider comprises of, firstly, a subthreshold
nondegenerate optical parametric amplifier (NOPA)
\cite{ou92a,ou92b,kimble92} for which the two external output
fields first pass through Faraday isolators and then each feed
into a different linearly damped optical cavity for which one
mirror is perfect and the other is lossy. The axes of both
cavities coincide with the $x$-axis. Each cavity supports an
electromagnetic field mode of frequency $\omega_{jc}$ that is
described by the annihilation operator $a_{j}$, where $j$
enumerates the cavities. Within the $j^{\rm th}$ cavity lie $N$
identical two-level atoms which each possess an internal
transition frequency of $\omega_{a}$. These are trapped in a
linear configuration parallel to the $x$-axis by, firstly, a
one-dimensional far-off-resonance dipole-force trap (FORT)
\cite{miller93,lee96}. This consists of cavity mode of frequency
$\omega_{T}$ which exhibits a standing-wave pattern along the
$x$-axis. The frequency $\omega_{T}$ is strongly detuned from all
atomic resonant frequencies and thus the cavity mode's field
exerts dipole forces on the atoms, trapping each of them near a
separate node of the field. In addition, the FORT's axis is
parallel to the $x$-axis and it thus traps the atoms in the $x$
direction. The atoms are also tightly confined in the $y$ and $z$
directions by a two-dimensional far-off resonance optical lattice
\cite{jessen96,haycock97,deutsch98} and thus move negligably in
these directions. The combined effect of all the trapping fields
is to confine each atom in its own one-dimensional trap parallel
to the $x$-axis. Furthermore, the atoms are located such that the
mean position of each atom coincides with a node of the cavity
field described by $a_{j}$. The annihilation operator
$b_{jx}^{(m)}$ describes the vibrational motion of the $m^{\rm
th}$ atom in the $j^{\rm th}$ trap in the $x$ direction. Finally,
external lasers of frequency $\omega_{L}$ whose beams are
perpendicular to the $x$-axis are incident on all atoms and thus
Fig.~\ref{ch3:fig_two} illustrates the system under consideration.

%
%Hamiltonian for a cavity
%

The Hamiltonian for the $j^{\rm th}$ optical cavity and the $N$
atoms within it is
\begin{eqnarray} \nonumber
H_{\rm j\:total} & = & H_{j0}^{\rm atom} + H_{j0} +
H_{jI} \\
& & + \kappa (a_{j} R^{\dag}_{j} + a^{\dag}_{j} R_{j}) + H_{\rm j\:res},
\end{eqnarray}
where $H_{j0}^{\rm atom}$ is the free Hamiltonian for the
vibrational states of the atoms and $H_{j0}$ is the free
Hamiltonian for the cavity field and the atoms' internal states.
The term $H_{jI}$ is the interaction Hamiltonian describing the
Raman processes involving the cavity field, the external lasers
and the atoms. To be more specific, $H_{j0}^{\rm atom} = \hbar
\nu_{jx} \sum_{m=1}^{N} \left( b_{jx}^{(m)\:\dag} b_{jx}^{(m)}
+\frac{1}{2}\right)$, where $\nu_{jx}$ is the vibrational
frequency of the FORT in the $j^{\rm th}$ cavity (which we call
the $j^{\rm th}$ FORT) in the $x$ direction. This frequency is
equal to $\omega_{jc} - \omega_{L}$. The Hamiltonian $H_{j0}$ is,
in a frame rotating at frequency $\omega_{L}$,
\begin{equation} \label{free_hamiltonian}
H_{j0} = \hbar \delta a_{j}^{\dag} a_{j} + \hbar \Delta \sum_{m=1}^{N} \sigma_{j+}^{(m)}
\sigma_{j-}^{(m)},
\end{equation}
where $\delta_{j} = \omega_{jc} - \omega_{L}$, $\Delta= \omega_{a}
-\omega_{L} $ and $\sigma_{j+}^{(m)}$ and $\sigma_{j-}^{(m)}$ are
raising and lower operators for the internal states of the $m^{\rm
th}$ atom in the $j^{\rm th}$ trap. The term $H_{jI}$ is
\begin{eqnarray} \nonumber
H_{jI}& = & \hbar \sum_{m=1}^{N} {\cal E}_{L}(y,z,t)
\sigma_{j+}^{(m)}+{\cal E}^{*}_{L}(y,z,t)
\sigma_{j-}^{(m)} \\
& & + \hbar g_{0} \sum_{m=1}^{N} \sin (k_{j} x_{jm}) \bigl(
a^{\dag}_{j} \sigma_{j-}^{(m)} +a_{j} \sigma_{j+}^{(m)} \bigr),
\end{eqnarray}
where ${\cal E}_{L}$ is the complex amplitude for all external
lasers, $k_{j}= \omega_{jc}/c$ and $g_{0}$ ($g_{0} \in
\mathbb{R}$) is the coupling constant for the atom-field
interaction. Finally, $H_{\rm j\:res}$ is the Hamiltonian for the
external reservoir that couples to the $j^{\rm th}$ cavity for
which $R_{j}$ is a reservoir annihilation operator and $\kappa$ is
a damping constant.

%
%asssumptions about model
%

The feasible assumptions below are made about the system in order to simplify
calculations for it and to focus on its most important aspects:
\begin{enumerate}
\item The cavity field and external laser frequencies are
appreciably detuned from $\omega_{a}$ and
all two-level ions are initially in the ground state.
Thus, the excited internal states
are sparsely populated and spontaneous emission effects are negligible
and can be ignored.
\item Vibrational decoherence occurs over a timescale much longer than that of the
interactions producing quantum state exchange and consequently can be ignored.
\item The wavelength of the cavity mode described by $a_{j}$ is much greater than
the distance that any atom in the $j^{\rm th}$ trap strays from
the cavity-field node about which it is trapped. Thus, $\sin
(kx_{jm}) \simeq k x_{jm} << 1$ and hence all atoms experience a
potential that is, to a good approximation, harmonic. This
justifies the form of $H_{j0}^{\rm atom}$.
\item All atoms are tightly confined in the $y$ and $z$
directions. This allows us to ignore the $y$ and $z$ dependence of
the external laser fields and thus, assuming ${\cal E}_{L}$ is
time independent, it follows that ${\cal E}_{L}(y,z,t) \simeq
{\cal E} e^{-i\phi_{L}}$, where ${\cal E}$ is a real
time-independent amplitude.
\item The damping parameter $\kappa$ is such that $\nu_{jx} \gg \kappa
\gg g_{0} \eta_{jx} \sqrt{N} {\cal E}/ \Delta$, where $\eta_{jx} =
k_{j} \sqrt{\hbar/2m \nu_{jx}}$, where $m$ is the mass of each
atom.
\end{enumerate}

%
%simplify H
%

Given the assumptions above, we can write $H_{\rm j\:total}$ in
terms of normal-mode creation and annihilation operators (by
adiabatically eliminating the evolution of the internal states) as
\begin{eqnarray} \nonumber
H_{\rm j\:total} & = & \hbar \sum_{m=1}^{N} \nu_{jx} \left(
B_{jx}^{\dag\:(m)} B_{jx}^{(m)}+\frac{1}{2} \right) + \hbar \delta
a^{\dag} a \\ \nonumber & - & \frac{\hbar N {\cal E}^{2}}{\Delta}
- \frac{\hbar g_{0}  \eta_{jx} \sqrt{N} {\cal E} }{\Delta} \times \\
\nonumber & &  \left( \sum_{m=1}^{N}
(B_{jx}^{(m)}+B_{jx}^{(m)\:\dag})
\bigl( a^{\dag}_{j}e^{-i\phi_{L}} +a_{j}  e^{i\phi_{L}}\bigr) \right) \\
& + & \kappa(a_{j} R^{\dag}_{j} + a^{\dag}_{j} R_{j}) + H_{\rm
j\:res}, \label{collective_modes}
\end{eqnarray}
where $B_{jx}^{(m)}$ is the annihilation operator for the $m^{\rm
th}$ normal mode for the $j^{\rm th}$ trap in the $x$ direction.
For example, $B_{jx}^{(1)}$ is a centre-of-mass mode annihilation
operator which is $B_{jx}^{(1)} = 1/\sqrt{N} \left( b_{jx}^{(1)} +
b_{jx}^{(2)} + \ldots b_{jx}^{(N)} \right)$ whilst $B_{jx}^{(2)}$
is the annihilation operator for the {\em breathing mode} which is
$B_{jx}^{(2)} = 1/\sqrt{2} \left( -b_{jx}^{(1)} +
b_{jx}^{(2)}\right)$ when $N=2$.

%
%by analogy qusex occurs into COM mode
%

Comparing Eqn~(\ref{collective_modes}) to
Eqn~(\ref{adiabatic_elim}), we see that $B_{jx}^{(1)}$ in
Eqn~(\ref{collective_modes}) plays an almost identical role to
that of $b_{x}$ in Eqn~(\ref{adiabatic_elim}). Given that
$\sin(kx) \simeq \eta_{x} (b_{x} + b_{x}^{\dag})$ in
Eqn~(\ref{adiabatic_elim}), the only difference between the forms
in which the two operators appear results from a factor of
$\sqrt{N}$ appearing in front of ${\cal E}$ in
Eqn~(\ref{collective_modes}). As a consequence, $B_{jx}^{(1)}$ in
Eqn~(\ref{collective_modes}) couples to the cavity mode $a_{j}$
identically
--- aside from the factor of $\sqrt{N}$ --- to the manner in which
$b_{x}$ couples to $a$. It follows that as quantum state exchange
takes place in the system described by $H_{\rm total}^{\rm
single}$ with information about an input electromagnetic field
being transferred to $b_{x}$, it also occurs in the system
described by $H_{\rm j\:total}$ due to the correspondence between
the two system's Hamiltonians. Thus, in the latter system,
information about the input field is transferred to the
centre-of-mass mode for the trapped atoms in the $x$ direction
{\em just as if this mode was a vibrational mode for a single
harmonically trapped atom}. The only difference between the
$N$-atom case and one described by $H_{\rm total}^{\rm single}$ is
that the effective coupling in the former case is increased by a
factor of $\sqrt{N}$. This conclusion can also be verified via
comparing the Langevin equations for $B_{jx}^{(m)}$ and $b_{x}$.
Due to symmetry considerations, collective modes other than the
centre-of-mass modes do not absorb any photons in modes $a_{1}$
and $a_{2}$. Thus, assuming these other modes are initially in
vacuum states, they remain so during quantum state exchange.

%
%\psi_{CM}
%

In \cite{parkins2}, it was shown that we can transfer the intracavity
steady-state for the subthreshold nondegenerate parametric amplifier which is
\begin{equation}
| \psi \rangle = \frac{1}{\cosh r}
\Sigma_{n=0}^{\infty} \tanh^{n} r
| n \rangle_{\rm 1} | n \rangle_{\rm 2},
\end{equation}
where the subscripts 1 and 2 denote the two output modes and $r$
is a real squeezing parameter, into the vibrational states in the
$x$ direction for two single trapped atoms in different harmonic
traps. Using the correspondence between the quantum state exchange
processes involving a single harmonically trapped atom and $N$
harmonically trapped atoms demonstrated above, it follows that in
the system illustrated in Fig.~\ref{ch3:fig_two} we can transfer
$| \psi \rangle$ into the centre-of-mass modes in the $x$
direction for the two sets of $N$ trapped atoms thus producing, in
the steady state,
\begin{equation}
| \psi_{\rm CM} \rangle = \frac{1}{\cosh r}
\Sigma_{{\cal N}=0}^{\infty} \tanh^{{\cal N}} r
| {\cal N} \rangle_{1} | {\cal N} \rangle_{2},
\end{equation}
where  $| {\cal N} \rangle_{j}$ denotes the {\em centre-of-mass}
vibrational number state for the $x$ direction with eigenvalue
${\cal N}$ for the atoms in the $j^{\rm th}$ FORT. In writing this
state we have omitted the states of collective modes other than
the centre-of-mass modes as we have assumed these other modes are
in vacuum states throughout the quantum state exchange process.

%
%practicality
%

Importantly, the process of creating $|\psi_{\rm CM} \rangle$ just
outlined does not seem to be overly experimentally infeasible.
This is so as optical cavities and nondegenerate optical
parametric amplifiers have been widely realized quantum optical
laboratories for some time.  In addition, neutral atoms have been
confined within standing-wave dipole-force traps that, in turn,
lie within optical cavities \cite{van_enk01}. Relatedly,
experiments in which a single harmonically trapped ion has been
placed within an optical cavity have been conducted
\cite{mundt02}.
\subsection{Summary} \label{summary}

%
%what we do in this paper
%

In this paper, we explore multipartite entanglement in relation to
quantum state exchange and in Section~\ref{section_qual}, follow
on from a multipartite entanglement condition implicit in work by
D\"{u}r and Cirac \cite{dur_cirac} by presenting a novel
condition. The satisfaction of this novel condition implies that
any pure state comprising of $M$ subsystems is $M$-way entangled.
Here, an $M$-way entangled state is one possessing entanglement
that spans across $M$ subsystems as does the generalized GHZ state
$| \psi \rangle = |0\rangle^{\otimes M}  +  |1\rangle^{\otimes
M}$. After presenting this condition, we then use it to show {\em
qualitatively} that $|\psi_{\rm CM} \rangle$ is $2N$-way
entangled. In Section~\ref{ch5:quant}, we {\em quantitatively}
consider the entanglement in $|\psi_{\rm CM} \rangle$. We
introduce a novel multipartite entanglement measure for pure
states we call the {\em entanglement of minimum bipartite entropy}
or $E_{\rm MBE}$ which is the minimum of the von Neumann entropies
of all the reduced density operators obtainable from some pure
state of interest by tracing over some of the subsystems for the
state. After this, we use $E_{\rm MBE}$ to calculate a {\em lower
bound} for the amount of 4-way, 6-way and 8-way entanglement in
$|\psi_{\rm CM} \rangle$ for $N=2,3,4$ respectively for a range of
$r$ values. Finally, we discuss the nature of our results.

%
%reasons for looking at entanglement in \psi_{CM}
%

It is interesting to investigate the nature $| \psi_{\rm CM}
\rangle$'s $2N$-way entanglement for a number of reasons. Firstly,
it has been claimed --- but not demonstrated --- $| \psi_{\rm CM}
\rangle $ is ``entangled state of all 2[$N$] \ldots atoms''
\cite{parkins1}. It is thus interesting to investigate $|
\psi_{\rm CM} \rangle$'s $2N$-way entanglement in order to see if
this implied claim is true. Secondly, it is interesting to
investigate $| \psi_{\rm CM} \rangle$'s $2N$-way entanglement as
it is a {\em massive-particle} state which is important as, to
date, mostly {\em massless} photons have been used to
experimentally investigate entanglement. Thirdly, if the claim is
true, then it means that $| \psi_{\rm CM} \rangle$ is a state
consisting of $2N$ entangled harmonic oscillators, each possessing
an {\em infinite-dimensional} Hilbert space as opposed to the
two-dimensional Hilbert space of a qubit, that is $2N$-way
entangled.

\section{Qualitative results} \label{section_qual}

\subsection{Negative partial transpose sufficient condition} \label{npt}

%
%background to NPT condition
%

Assume that, for a certain state $\rho$, we wish to know the answer to
the question ``Does $\rho$ contain at least some $M$-way entanglement?''
Whilst answering this question does not tell us everything about the
nature of $\rho$'s $M$-way entanglement, it nevertheless tells us
something of interest.
One way to answer it, provided that $\rho$ consists of qubits, is to
use a condition that can be readily derived from work by D\"{u}r and Cirac
\cite{dur_cirac}. This condition involves negative partial transposes (NPTs)
\cite{peres96,horodecki96,horodecki97} and
thus we name it the {\em NPT sufficient
condition}. It is sufficient for the presence of
$M$-way entanglement for all $\rho$'s consisting of $P$ qubits, where $P \ge M$,
and is based on generalizing the notions of separability and inseparability to many-qubit systems.
Before stating the condition, it is first useful to mention two
things. Firstly, we define an {\em M-partite split} of $\rho$
\cite{dur_cirac}
to be a
division or split of $\rho$ into $M$ parts which each consist of one or more subsystems.
Secondly, we observe that $\rho$ can always be converted to a
state that is diagonal
in a certain basis by a ``depolarization'' process consisting of
particular local operations \cite{dur_cirac}. This basis consists of $M$-qubit
generalized GHZ states of the form
$|\psi \rangle =1/\sqrt{2} \left( |j\rangle |0\rangle \pm |2^{N-1}-j-1\rangle| 1 \rangle
\right)$, where $j$ is a natural number that we write in binary as
$M-1$ bits, i.e. $j \equiv j_{1} j_{2} \ldots j_{M-1}$, where $j_{x}$
is the $x^{\rm th}$ bit in $j$'s binary representation.
Given these two things, the NPT sufficient condition states that $\rho$
is $M$-way entangled for a given $M$-partite split if
the diagonal state that it depolarizes to is such that all
{\em bipartite splits} that {\em contain} the $M$-partite split have negative partial transposes.
A {\em bipartite split} is one that divides a system into two parts, i.e.
a $2$-partite split. Also, a bipartite split that {\em contains} an $M$-partite split
is one that does not separate members of any of the $M$ subsystems
onto two different sides of the bipartite split. That is,
one that does not cross any of the divisions created by the $M$-partite split.

\subsection{Result} \label{ch5:qual}

%
%intro. to subsection
%

Following on from the NPT sufficient condition, we propose a
necessary and sufficient condition for the existence of $M$-way entanglement
for $M$-partite pure states. Our condition is based on the traces of the squares of
reduced density matrices
obtained by tracing over some of the subsystems constituting
our system of interest. After
formulating it, we then use it to demonstrate that $| \psi_{\rm CM} \rangle$ contains
{\em some} $2N$-way entanglement. Our motivations for employing our
condition instead of the NPT sufficient condition are that i) it seems to be mathematically
simpler to calculate whether or not our condition is satisfied and
ii) as we are concerned with a pure state, our condition is stronger
than the NPT sufficient condition in
the sense that it is both necessary and sufficient as opposed to just being sufficient.

%
%present the condition
%

Our $M$-way entanglement condition utilizes the fact that when a
{\em pure} state $| \psi \rangle$ for $M$ subsystems is $M$-way
entangled then we {\em cannot} write it as $| \psi \rangle =|
\phi_{1} \rangle_{Q_{j}} \otimes | \phi_{2}
\rangle_{\bar{Q}_{j}}$, where $| \phi_{1} \rangle_{Q_{j}}$ and $|
\phi_{2} \rangle_{\bar{Q}_{j}}$ are the states for the subsystems
denoted by $Q_{j}$ and ${\bar{Q}_{j}}$ respectively and both
$Q_{j}$ and ${\bar{Q}_{j}}$ denote at least one subsystem. To put
this another way, when $| \psi \rangle$ is $M$-way entangled then
there is {\em no} way to represent it as the tensor product of two
pure states. Consequently, excluding all such possibilities
suffices to show, and is also, in general, necessary to show, that
$| \psi \rangle$ is $M$-way entangled. This can be done by first
checking that no single-subsystem state can be factored out from
the state of the remaining $M-1$ subsystems. We do this by
checking that the traces of the squares of all the reduced density
operators obtainable from $| \psi \rangle$ by tracing over one
subsystem are less than one. That is, that $Tr([\rho_{Q_{j}}]^{2})
< 1$, where $\rho_{Q_{j}}$ is the reduced density operator
obtained from $| \psi \rangle \langle \psi |$ by tracing over the
subsystem denoted by $Q_{j}$ for all $Q_{j}$ denoting just one
subsystem. We can then repeat this procedure, considering all
$Q_{j}$'s corresponding to all pairs of subsystems, then all
triples and so forth until we have considered all $Q_{j}$'s
corresponding to all sets of $R$ subsystems, where $R = \lfloor
M/2 \rfloor$, where $\lfloor x \rfloor$ is the largest integer
less than or equal to $x$. It is sufficient to only consider sets
of up to those corresponding to $\lfloor M/2 \rfloor$ subsystems
as a necessary condition for being able to factor out any larger a
number of subsystems from $| \psi \rangle$ is the ability to also
factor out $\lfloor M/2 \rfloor$ or fewer subsystems. Underlying
the process just described is that of seeing whether or not we can
exclude all the ways that $| \psi \rangle$ could fail to be
$M$-way entangled.

%
%formalized condition
%

Our condition can be formalized as {\bf Definition 1} which is as follows:
\begin{quote}
{\bf Definition 1} : For a pure state $| \psi \rangle$ for $M$ subsystems,
consider the set $Q$ whose members $Q_{j}$ are themselves sets of
subsystems for the system corresponding to $| \psi \rangle$. This set
$Q$ contains all sets of $P$ subsystems for
this system, where $1 \leq P \leq \lfloor M/2 \rfloor$.
Given this, $| \psi \rangle$ is $M$-way entangled iff, for all
$Q_{j}$, $Tr\left( [\rho_{Q_{j}}]^{2} \right) < 1$, where
$\rho_{Q_{j}}$ is the reduced density operator obtained by beginning with
$|\psi \rangle \langle \psi |$ and tracing over the subsystems $Q_{j}$.
\end{quote}
To illustrate {\bf Definition 1} consider, for example, the GHZ state
$|\psi\rangle_{\rm GHZ}=1/\sqrt{2}\left(|000\rangle_{123} + |111\rangle_{123} \right)$, where
the subscripts 1, 2 and 3 denote
subsystems of $| \psi \rangle_{\rm GHZ}$.
The parameter $P=\lfloor 3/2 \rfloor = 1$ and consequently
the set $Q$ comprises of all sets of one
subsystem and thus $Q=\{ \{1\}, \{2\}, \{3\} \}$,
where the numbers again denote subsystems for $|\psi\rangle_{\rm GHZ}$.
For the element $\{1\}$, for example, $Tr([\rho_{\{1\}}]^{2}) = 1/2$. Calculating
$Tr([\rho_{Q_{j}}]^{2})$ for all of $Q$'s other elements,
we find that it is 1/2 in all three cases. Thus, $| \psi \rangle_{\rm GHZ}$ satisfies {\bf Definition 1} and hence is
said to be 3-way entangled, as is the case.

%
%examples of condition A
%

To further explain {\bf Definition 1}, we now apply it to
determining whether the following four-party states are 4-way entangled :
\begin{enumerate}
\item\hspace*{-0.15cm}) $\;| \psi_{4}^{(1)} \rangle = 1/\sqrt{2} \bigl(
|0000\rangle_{1234} + |1111\rangle_{1234} \bigr)$,
\item\hspace*{-0.15cm})   $\;| \psi_{4}^{(2)} \rangle = 1/\sqrt{2} | 0 \rangle_{1} \otimes (|000\rangle_{234} + |111\rangle_{234})$
\item\hspace*{-0.15cm})  $\;| \psi_{4}^{(3)} \rangle = |\phi^{+} \rangle_{12}
\otimes | \phi^{+} \rangle_{34}$.
\end{enumerate}
Turning to 1.), we see that upon tracing over any single
subsystem, we produce a reduced density operator of the form
$\rho_{Q_{j}} = 1/2 (|000\rangle \langle 000| + |111\rangle
\langle 111|)$ for which $Tr([\rho_{Q_{j}}]^{2}) = 1/2$.
Similarly, tracing over any two subsystems produces a density
operator of the form $\rho_{Q_{j}}=1/2 (|00\rangle \langle 00| +
|11\rangle \langle 11|)$ for which, again,
$Tr([\rho_{Q_{j}}]^{2})=1/2$. Thus, {\bf Definition 1} gives the
correct result that $| \psi_{4}^{(1)} \rangle$ is 4-way entangled.
For 2.), tracing over the first subsystem produces
$|\psi_{4}^{(3)} \rangle = 1/\sqrt{2} \bigl( |000\rangle_{234} +
|111\rangle_{234} \bigr)$, which is a pure state and hence
$Tr([\rho_{Q_{j}}]^{2}) =1$ for the corresponding $j$.
Consequently, {\bf Definition 1} tells us that $|
\psi_{4}^{(2)}\rangle$ is {\em not} $4$-way entangled, as is the
case. For 3.), tracing over any one subsystem produces the mixed
state $\rho = I/2 \otimes |\phi^{+} \rangle_{34} \langle
\phi^{+}|$ and so we might be tempted to infer that  $|
\psi_{4}^{(3)} \rangle$ is 4-way entangled. However, when we trace
over subsystems 1 and 2 or subsystems 3 and 4 we produce the pure
state $|\phi^{+}\rangle$ for which  $Tr \left( [|\phi^{+}\rangle
\langle \phi^{+} |]^{2} \right)= 1$. Hence, {\bf Definition 1}
correctly tells us that $| \psi_{4}^{(3)} \rangle$ is not 4-way
entangled.

%
%switch to individual states
%

In applying {\bf Definition 1} to $| \psi_{\rm CM} \rangle$, we
first write $| \psi_{\rm CM} \rangle$ in terms of vibrational
number states for the $2N$ atoms involved as we wish to see if
they are $2N$-way entangled. As a step towards doing so, upon
observing that $| {\cal N} \rangle_{j} = \left( \left(
B^{(1)\:\dag}_{jx} \right)^{\cal N}/\sqrt{{\cal N}!} \right) | 0
\rangle_{j}$, we express $| {\cal N} \rangle_{j}$ in terms of
vibrational number states in the $x$ direction for {\em
individual} atoms as
\begin{eqnarray} \label{com_to_individual}
|{\cal N} \rangle_{j} =
\sum_{{\rm a.c.}(\vec{n},{\cal N})}
c(\vec{n},{\cal N}) | \vec{n} \rangle_{j},
\end{eqnarray}
where $\vec{n}$ is the $N$-component vector $(n_{1}, n_{2} \ldots
n_{N})$, the state $| \vec{n} \rangle_{j} = |n_{1} \rangle_{j}
\otimes |n_{2} \rangle_{j} \ldots |n_{N} \rangle_{j}$, where $|
n_{k} \rangle_{j}$ denotes a number state for the $k^{\rm th}$
atom in the $j^{\rm th}$ FORT and
\begin{eqnarray} \nonumber
c(\vec{n},{\cal N}) & = & _{j}\langle \vec{n} | {\cal N}\rangle_{j} \\ \nonumber
& = & \frac{
\left( \!\!
\begin{array}{c}
{\cal N} \\
n_{1}
\end{array}
\!\! \right)
\left( \!\!
\begin{array}{c}
{\cal N} -n_{1} \\
n_{2}
\end{array}
\!\! \right)
...
\left( \!\!
\begin{array}{c}
{\cal N} - n_{1}  ... - n_{N - 2} \\
n_{N-1}
\end{array}
\!\! \right)}
{\sqrt{{\cal N}! \times N^{{\cal N}}} } \\
& & \times \sqrt{n_{1}! n_{2}! ... n_{N}!}
\end{eqnarray}
The sum $\sum_{{\rm a.c.}(\vec{n},N)}$ denotes the sum over {\em
all combinations} of $n_{1}, n_{2} \ldots n_{N}$ such that
$\sum_{j=1}^{N} n_{j}=N$ \cite{rice94}. Using
Eqn~(\ref{com_to_individual}) to represent $|\psi_{\rm CM}\rangle$
in terms of vibrational number states for individual atoms, we
obtain
\begin{eqnarray} \label{individual} \nonumber
| \psi_{\rm CM} \rangle & = &
\frac{1}{\cosh r} \sum_{{\cal N}=0}^{\infty} \tanh^{{\cal N}} r
\left( \sum_{a.c.(\vec{n},{\cal N})}
c(\vec{n},{\cal N}) | \vec{n} \rangle_{1} \right) \\
& & \otimes  \left(  \sum_{a.c.(\vec{m},{\cal N})}
c(\vec{m},{\cal N}) | \vec{m} \rangle_{2}  \right).
\end{eqnarray}

%
%the qualitative result
%

We now show that the right-hand side of Eqn~(\ref{individual})
satisfies {\bf Definition 1} and thus that $|\psi_{\rm CM}
\rangle$ is $2N$-way entangled. We do this by first writing
$|\psi_{\rm CM} \rangle$ as the most general bipartite state
possible involving vibrational number states for individual atoms.
Next, we show that, upon tracing over the atoms in the half of the
bipartite split containing the lesser number of atoms and then
finding the trace of the square of the resulting reduced density
operator, that this is less than one. It follows that, for all
$j$, $Tr([\rho_{Q_{j}}]^{2})<1$. Hence, we satisfy {\bf Definition
1} and so $|\psi_{\rm CM} \rangle$ is $2N$-way entangled.

%
%actually do qualitative calc.
%

Dividing the atoms in $|\psi_{\rm CM}\rangle$ into subsystems $A$
and $B$ containing, respectively, $R$ and $2N-R$ atoms ($R \neq
0$), we can write $|\psi_{\rm CM}\rangle$ as
\begin{equation} \label{form}
|\psi_{\rm CM} \rangle
= \sum_{i=0}^{\infty}
c_{i} | f_{i} \rangle_{A} \otimes | g_{i} \rangle_{B},
\end{equation}
where $\| |f_{i}\rangle_{A} \| =\| |g_{i}\rangle_{B} \|=1$ and the
$|f_{i}\rangle_{A}$, but not necessarily the $|g_{i}\rangle_{B}$,
are mutually orthogonal. (As we can always write $|\psi_{\rm CM}
\rangle$ in biorthogonal form \cite{peres93}, there exist $|g_{i}
\rangle_{B}$ that {\em are} mutually orthogonal. However, we are
not concerned with this form in the current calculation and so do
not consider such a decomposition of $| \psi_{\rm CM} \rangle$.)
To give an example, when $N=2$ and $A$ contains the first atom in
the first trap
\begin{eqnarray} \nonumber
| \psi_{\rm CM} \rangle & = &  \frac{1}{\cosh r} |0 \rangle_{A} \otimes \left( |000 \rangle_{B}
+ \frac{\tanh r}{2}|101\rangle_{B} \right. \\ \nonumber
& & \left. +  \frac{\tanh r}{2}|110\rangle_{B}
+ \frac{\tanh^{2} r}{4} |202 \rangle_{B}  \right. \\ \nonumber
& & \left. + \frac{\sqrt{2} \tanh^{2}
r}{4} |211 \rangle_{B}   + \frac{\tanh^{2} r}{4}
|220 \rangle_{B}  + \ldots \right) \\ \nonumber
&  & + \frac{1}{\cosh r} |1 \rangle_{A} \otimes
\left( \frac{\tanh r}{2} |001 \rangle_{B} \right. \\ \nonumber
& & \left. +  \frac{\tanh r}{2} |010\rangle_{B} + \frac{\tanh^{2}
r}{\sqrt{8}} |102 \rangle_{B} \right. \\ \nonumber
& & \left. + \frac{\sqrt{2} \tanh^{2} r}{2} |111 \rangle_{B}
+ \frac{\tanh^{2} r}{2} |120 \rangle_{B} + \ldots \right) \\ \nonumber
&  & + \frac{1}{\cosh r} |2 \rangle_{A} \otimes
\left( \frac{\tanh^{2} r}{4} |002 \rangle_{B} \right. \\ \nonumber
& & \left. + \frac{\sqrt{2} \tanh^{2} r}{8} |011 \rangle_{B}
+ \frac{\tanh^{2} r}{4} |020 \rangle_{B} + \ldots \right) \\
& & +   \ldots \label{com_to_individ},
\end{eqnarray}
where $|x\rangle_{A} = |n_{1}=x\rangle_{A}$
and $| x_{1} x_{2} x_{3}\rangle_{B} =|n_{2} =x_{1}, m_{1}=x_{2}, m_{2}=x_{3} \rangle_{B}$.
Here, for example, $c_{0}=1/\cosh r$,
$| f_{0} \rangle_{A} = |0\rangle_{A}$,
$c_{1} = 1/\cosh r$,
$| f_{1} \rangle_{A} = |1\rangle_{A}$,
\begin{eqnarray} \nonumber
|g_{0} \rangle_{B} & = & \frac{1}{\sqrt{{\cal M}_{0}}} \left(
|000 \rangle_{B}
+ \frac{\tanh r}{2}|101\rangle_{B} \right. \\ \nonumber
& & \left. + \frac{\tanh r}{2} |110\rangle_{B}
|\rangle  + \frac{\tanh^{2} r}{4} |202 \rangle_{B} \right. \\ \nonumber
& & \left. + \frac{\sqrt{2} \tanh^{2} r}{4} |211 \rangle_{B}
+ \frac{\tanh^{2} r}{4} |220 \rangle_{B} + \ldots \right) \\ \nonumber
& {\rm and} &  \\ \nonumber
|g_{1} \rangle_{B} & = &
\frac{1}{\sqrt{{\cal M}_{1}}} \left(
\frac{\tanh r}{2} |001 \rangle_{B} + \frac{\tanh r}{2} |010\rangle_{B}
\right.  \\ \nonumber
& & \left. + \frac{\tanh^{2} r}{\sqrt{8}} |102 \rangle_{B}
+ \frac{\sqrt{2} \tanh^{2} r}{2} |111 \rangle_{B} \right. \\ \nonumber
& & \left. + \frac{\tanh^{2} r}{2} |120 \rangle_{B} + \ldots \right),
\end{eqnarray}
where ${\cal M}_{0}$ and ${\cal M}_{1}$ normalize
$|g_{0}\rangle_{B}$ and $|g_{1}\rangle_{B}$.
Upon tracing over $A$ in Eqn~(\ref{form})
and squaring the resulting reduced density operator $\rho_{Q_{A}}$, we obtain
\begin{equation} \label{rho_squared}
[\rho_{Q_{A}}]^{2} = \sum_{i,j = 0}^{\infty,\infty} c_{i}^{2}  c_{j}^{2}
|g_{i} \rangle \langle  g_{i} |  g_{j} \rangle \langle g_{j} |.
\end{equation}
Calculating the trace of $[\rho_{Q_{A}}]^{2}$ yields
\begin{equation} \label{dees}
Tr\left( [\rho_{Q_{A}}]^{2} \right)
=\sum_{i,j=0}^{\infty,\infty} c_{i}^{2}c_{j}^{2} |d_{ij}|^{2},
\end{equation}
where $d_{ij}=\langle g_{i}|g_{j} \rangle$. As the trace of a
density operator is always one, we know that
\begin{equation}
\sum_{i,j=0}^{\infty,\infty} c_{i}^{2} c_{j}^{2}= \left(\sum_{i=0}^{\infty}
c_{i}^{2} \right) \times \left( \sum_{j=0}^{\infty} c_{j}^{2}\right) = 1.
\end{equation}
It thus follows from Eqn~(\ref{dees}) that, as
$c_{i} \neq 0$ for all $i$, if $|d_{ij}|^{2}<1$ for at least one $d_{ij}$ then
$Tr\left( [\rho_{Q_{A}}]^{2} \right)<1$.

%
%showing d_{0,1} <1
%

As the centre-of-mass state $|{\cal N}\rangle_{1} |{\cal
N}\rangle_{2}$ has an even number of centre-of-mass phonons in
total ($2{\cal N}$), when we express it as a sum of vibrational
number states for individual atoms, these states all contain an
even number of individual phonons in total. Furthermore, because
$|\psi_{\rm CM}\rangle$ contains the state $|{\cal N}=0
\rangle_{1} | {\cal N}= 0 \rangle_{2}$, one $|f_{i}\rangle_{A}$ in
Eqn~(\ref{individual}), which we denote by $|f_{i}^{\rm
zero}\rangle_{A}$, is a tensor product of ground states for some
of the $2N$ atoms in $|\psi_{\rm CM}\rangle$. For example, in
Eqn~(\ref{com_to_individ}), $| f_{i}^{\rm zero} \rangle_{A} =| 0
\rangle_{A}$. Given that, in general, $| f_{i}^{\rm zero}
\rangle_{A}$ contains zero individual phonons, only states with an
{\em even} number of individual phonons in total are present in
the $| g_{i} \rangle_{B}$ with the same index $i$, which we denote
by $|g_{i}^{\rm zero} \rangle_{B}$. This is so as we require the
total number of individual phonons in $| f_{i}^{\rm zero}
\rangle_{A} \otimes |g_{i}^{\rm zero} \rangle_{B}$ to be even.

%
%f=odd term
%

In addition to $| f_{i}^{\rm zero} \rangle_{A}$, because
$|\psi_{\rm CM}\rangle$ includes the term $|{\cal N}=1 \rangle_{1}
| {\cal N}=1 \rangle_{2}$, there also exists an $| f_{i}
\rangle_{A}$ in Eqn~(\ref{individual}) containing just one
individual phonon which we denote as $|f_{1}^{\rm one}
\rangle_{A}$. For example, in Eqn~(\ref{com_to_individ})
$|f_{1}^{\rm one} \rangle_{A}=|1\rangle_{A}$. In general, the
$|g_{i}\rangle_{B}$ with the same index $i$ as $| f_{i}^{\rm one}
\rangle_{A}$, which we denote by $|g_{i}^{\rm one}\rangle_{B}$,
comprises of states with an {\em odd} number of individual phonons
in total as dictated by the requirement that the total number of
individual phonons for $| f_{i}^{\rm one} \rangle_{A} \otimes
|g_{i}^{\rm one} \rangle_{B}$ is even. Thus, $|g_{i}^{\rm
one}\rangle_{B}$ is orthogonal to $|g_{i}^{\rm zero}\rangle_{B}$
and the corresponding $|d_{ij}|^{2}=|\langle  g_{i}^{\rm zero}
|g_{i}^{\rm one} \rangle|^{2}=0$. Returning to the right-hand side
of Eqn~(\ref{dees}), this means that $Tr\left( [\rho_{Q_{A}}]^{2}
\right) < 1$ for $Q_{A}$ and thus that {\bf Definition 1} is
satisfied. This allows us to infer that $| \psi_{\rm CM} \rangle$
is $2N$-way entangled and consequently we have verified the
assertion that $| \psi_{\rm CM} \rangle$ is an ``entangled state
of all 2[N] \ldots atoms'' --- except, of course, when $r=0$.

\section{Quantifying the amount of $2N$-way entanglement in
$|\psi_{\rm CM} \rangle$} \label{ch5:quant}

\subsection{Theory} \label{quant_theory}

In the previous subsection we presented a qualitative result
which showed that $|\psi_{\rm CM} \rangle$ possessed some
$2N$-way entanglement. However, we would also like to know
how much $2N$-way entanglement $|\psi_{\rm CM} \rangle$ contains.
For this reason, we present
an novel {\em quantitative} measure of $M$-way entanglement
for $M$-partite {\em pure states}, for arbitrary $M$.
This measure is
based on the von Neumann entropies of reduced density operators
produced by considering all bipartite splits for some state of interest.
We call it the {\em entanglement of
minimum bipartite entropy} or $E_{\rm MBE}$, which we
soon define. After this, we then argue that it is a plausible measure
and finally use it to calculate a {\em lower bound} on the amount
of $2N$-way entanglement in $|\psi_{\rm CM} \rangle$.

%
%our quantification
%

For a pure state $| \psi \rangle$ with $M$ subsystems, $E_{\rm MBE}$ is
\begin{equation} \label{e_mbe}
E_{\rm MBE}(| \psi \rangle)={\rm min} ({\bf S_{\rm all}}),
\end{equation}
where ${\bf S_{\rm all}}$ is the set containing
the von Neumann entropies of {\em all} the reduced density operators
obtained from $|\psi\rangle \langle \psi |$
by tracing over a set of $P$ subsystems in $|\psi\rangle$, where $1 \leq P \leq \lfloor M/2 \rfloor$.
The function min($X$) returns the smallest element of the set $X$.
Thus, as the von Neumann entropies of both sides of
any bipartite split of $| \psi \rangle$ are equal \cite{nielsen00}
(p. 513), ${\bf S_{\rm all}}$
contains the von Neumann entropies for {\em all} the
reduced states that we can generate from $| \psi \rangle$. For example, when
$| \psi \rangle= |\psi \rangle_{\rm GHZ}= 1/\sqrt{2}\left( |000\rangle_{123}+|111\rangle_{123}\right)$,
the sets of subsystems containing $P$ members
that we trace over in obtaining
${\bf S_{\rm all}}$ are
$\{ 1 \}$, $\{ 2\}$ and $\{ 3\}$,
where the numbers denote either the `1', `2' or `3' subsystems of $|
\psi \rangle_{\rm GHZ}$.
As the von Neumann entropy of the state $\rho$ is $S(\rho) = -Tr(\rho
\log_{2} \rho)$ \cite{nielsen00,preskill98},
the von Neumann entropy for the reduced density operator generated
from $|\psi \rangle_{\rm GHZ} \langle \psi |$ upon tracing over the subsystem
denoted by any one of these sets is 1. Hence ${\bf S_{\rm all}}=\{1,1,1\}$
and so $E_{\rm MBE}(|\psi\rangle_{\rm GHZ}) = 1$. We thus
we say that $|\psi\rangle_{\rm GHZ}$ has 1 unit of 3-way entanglement.

%
%distance measure
%

\begin{figure}[ht]
\center{\epsfig{figure=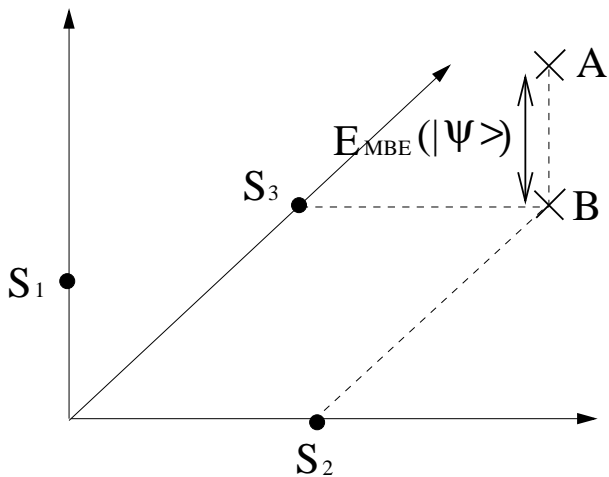,width=80mm}}
\caption{\label{picture} Co-ordinate space illustrating that
$E_{\rm MBE}$ can be seen as a distance-based entanglement
measure. The distance between $A$ (the point representing ${\bf
S_{\rm all}} ( | \phi \rangle ) = \{S_{1},S_{2},S_{3}\}$, where
$|\phi\rangle$ is $3$-way entangled and $S_{1} < S_{2},S_{3}$) and
$B$ (the point representing ${\bf S_{\rm all}} ( | \phi_{\rm zero}
\rangle ) = \{0, S_{2}, S_{3}\}$, where $| \phi_{\rm zero}
\rangle$ appears to be the closest pure state with no $3$-way
entanglement to $| \phi \rangle$) is $E_{\rm MBE}$. The quantities
$S_{1}$, $S_{2}$ and $S_{3}$ are dimensionless.}
\end{figure}

To provide some insight into $E_{\rm MBE}$, it is now shown that it can be thought of as
a {\em distance-based} measure of $M$-way entanglement. That is, as measuring the distance between $|\psi\rangle$ and the closest pure
state with zero $M$-way entanglement given a certain metric. To understand this, observe
that, naively, it seems reasonable to think that there exists a pure state $|\psi_{\rm zero}\rangle$
with zero $M$-way entanglement that has an identical ${\bf S_{\rm all}}$
to $|\psi\rangle$'s except for one element. This element
corresponds to the smallest element of ${\bf S_{\rm all}} (|\psi\rangle)$ and is zero.
The next step in comprehending the distance-based nature of $E_{\rm MBE}$ is representing
${\bf S_{\rm all}} (| \psi \rangle)$ and ${\bf S_{\rm all}}  (| \psi_{\rm zero} \rangle)$
by points $A$ and $B$ respectively in a co-ordinate space for which
each co-ordinate denotes the possible values of an element of
either ${\bf S_{\rm all}}(|\psi\rangle)$ or  ${\bf S_{\rm all}}(|\psi_{\rm zero}\rangle)$.
That is, a space that graphically represents ${\bf S_{\rm all}}(|\psi\rangle)$ and ${\bf S_{\rm all}}(|\psi_{\rm zero}\rangle)$.
For such a space, we observe that no pure state
with zero $M$-way entanglement is represented by a point closer
to $A$ than $B$. It is in this sense that we think of
$| \psi_{\rm zero} \rangle$ as being the closest pure state to $|\psi\rangle$ with zero
$M$-way entanglement. Finally, the distance-based nature
of $E_{\rm MBE}(| \psi \rangle)$ can be seen by observing that the distance between $A$ and $B$ is $E_{\rm MBE}(| \psi \rangle)$.
This point is illustrated in Fig.~\ref{picture} for the 3-way entangled state $| \phi \rangle$ comprising of three subsystems for
which ${\bf S_{\rm all}}(| \phi \rangle)=\{S_{1},S_{2},S_{3}\}$,
where $S_{1} < S_{2},S_{3}$ and $S_{1},S_{2},S_{3} \neq 0$. Naively, the closest pure state
to $|\phi\rangle$ with no 3-way entanglement $| \phi_{\rm zero} \rangle$
seems to be such that ${\bf S_{\rm all}}(| \phi_{\rm zero} \rangle )  = \{0, S_{2}, S_{3}\}$.
Representing ${\bf S_{\rm all}}(| \phi \rangle)$ and
${\bf S_{\rm all}}(| \phi_{\rm zero} \rangle )$ graphically in the manner described above
by points A and B in Fig.~\ref{picture}, we observe that the distance between them is $S_{1}=E_{\rm MBE}$.
Generalizing this notion, we see that $E_{\rm MBE}$ can be viewed as measuring the distance between
$|\psi\rangle$ and the nearest pure state with zero $M$-way
entanglement. This distance seems to be a plausible measure of $| \psi \rangle$'s
$M$-way entanglement and thus $E_{\rm MBE}$ appears to have an underlying intuitive motivation.

%
%analogy for entanglement of MBE
%

To further highlight the plausibility of $E_{\rm MBE}$, consider the following analogy.
Imagine an ordinary chain with $M$ links.
If $M-1$ of these are strong and the other one
is weak, then the chain is close to breaking and
so only has a small amount of ``nonbroken-ness'' --- even
though all but one of the links
are solid. This is so as nonbroken-ness is a wholistic property
that is a manifestation of the nature of all $M$ links.
Relating this to $E_{\rm MBE}$, just
as nonbroken-ness is a wholistic property, so
$E_{\rm MBE}$ measures a wholistic property, namely $M$-way entanglement,
that relates to the nature of
all $M$ subsystems of $M$-partite states.
In analogy with a chain with just one weak link, an $M$-partite pure
state for which all members of ${\bf S_{\rm all}}$ are large, except for one,
is very close to possessing no $M$-way entanglement.
In this way, we see that $E_{\rm MBE}$ and, in particular,
the presence of the min function in it seem plausible.

%
%desiderata
%

Another positive feature of $E_{\rm MBE}$
is that it satisfies three well-known desiderata
for {\em bipartite} entanglement measures \cite{plenio98}, as we now show.
(It seems plausible that these should also be
desiderata for {\em multipartite} entanglement measures.)
They are :
\begin{enumerate}
\item\hspace*{-0.15cm}) The proposed entanglement measure is zero for all product states.
\item\hspace*{-0.15cm}) The proposed entanglement measure is invariant under local unitaries.
\item\hspace*{-0.15cm}) The proposed entanglement measure does not
increase on average under local operations, classical communication (LOCC)
and division into subensembles.
\end{enumerate}
Beginning with 1.), if the state of interest is a product state,
where we define a product state to be
one for which we can factor out the state of
at least one of the subsystems, then
at least one member of ${\bf S_{\rm all}}$ is zero and so $E_{\rm MBE}$
is also zero, as we desire.
Turning to 2.), we note that for a general bipartite split, the von
Neumann entropy of the
reduced density matrix obtained by tracing over
the subsystems on the side of the split
with the lesser number of particles
is invariant under unitary transformations which act on only one subsystem.
Consequently, if we define local unitaries to be those which act just on
a single subsystem, then $E_{\rm MBE}$ satisfies 2.)

%
%condition 3.)
%

In considering 3.), it is important to remember that
$E_{\rm MBE}$ is only for pure states and thus we ignore local operations
that convert $| \psi \rangle$ to a mixed state. For example, we do not consider local operations that transform
$| \psi \rangle$ to a state that is close to a maximally mixed state
and thus has large values for the von Neumann entropies of all its reduced states.
We choose this example as such local operations increase the value of ${\rm min}({\bf S_{\rm all}})$
for a system of interest. However, they manifestly do not increase
its $M$-way entanglement but instead transform its state into one for which $E_{\rm MBE}$ is not applicable.
With this constraint in mind,
we define a local operation to be one that involves just one
subsystem, such
as a projective measurement on a single subsystem. Given this definition, it can be shown that for
{\em bipartite} pure states LOCC
and division into subensembles cannot increase the average entanglement
of any state as measured by the von Neumann entropy of its reduced
states (entropy of entanglement) \cite{plenio98}.
It follows that they also cannot increase any
member of ${\bf S_{\rm all}}(|\psi\rangle)$, on average, as these faithfully
measure the bipartite entanglement in $| \psi \rangle$ given some bipartite split for it.
Thus, $E_{\rm MBE}$ also cannot increase, on average, under LOCC
and division into subensembles and so $E_{\rm MBE}$
satisfies 3.)

%
%additivity
%

Another well-known desideratum for a bipartite entanglement measure is that
it is {\em additive over tensor products} \cite{plenio98}.
However, it can be shown that $E_{\rm MBE}$ is {\em superadditive}.
That is, that the $M$-way entanglement of a combined state generated from
two states with $a$ and $b$ units of $M$-way entanglement can
be greater than $a + b$ (but, importantly, not when $M=2$).
It is an open question as to whether or not multipartite entanglement is
additive and so we do not know if the superadditivity of $E_{\rm MBE}$ represents a flaw.

%
%making sure E_{\rm MBE} reduces to bipartite EOF
%

For $E_{\rm MBE}$ to be a reasonable measure, it ought to reduce to the standard pure state
bipartite entanglement measure of the entropy of entanglement.
For $E_{\rm MBE}$, when $N=1$, we have
$E_{\rm MBE}= {\rm min}({\bf S_{\rm all}}) = S_{\{1\}}$,
where $S_{\{1\}}$ is the von Neumann entropy for
the reduced density operator $\rho_{Q_{\{1\}}} =
{\rm Tr_{1}}( | \psi \rangle \langle \psi |)$,
and so we recover the desired measure, namely the entropy of entanglement.
Finally, $E_{\rm MBE}$ seems to be plausible as
for $|\psi\rangle = {\sqrt c} |0\rangle^{\otimes N}
+ \sqrt{1-c}|1\rangle^{\otimes N}$, where $c \in [0,1]$ and $N$ is a positive integer,
$E_{\rm MBE} = -c \log_{2} c - (1-c) \log_{2}(1-c)$.
This expression increases monotonically in the interval $c \in [0,1/2]$
and attains its maximum value of one for $c=1/2$. Such behaviour seems reasonable.

\subsection{Results} \label{quant_results}

In this subsection we use $E_{\rm MBE}$ to calculate {\em lower
bounds} on the amount of $2N$-way entanglement present in
$|\psi_{\rm CM} \rangle$ for $N=2,3,4$, for a range of $r$ values.
We obtain these lower bounds by, first, calculating
$Tr([\rho_{Q_{j}}]^{2})$ for a general $Q_{j}$. Next, we determine
the {\em linear entropy} $S_{L} (\rho_{Q_{j}})$ \cite{bose00} from
the relation $S_{L}( \rho_{Q_{j}}  ) = 1-Tr([\rho_{Q_{j}}]^{2})$
and then use the fact that $S_{L}(\rho) / \log_{2} {\rm e} \leq
S(\rho)$ to obtain our lower bounds. We calculate a lower bound
rather than $E_{\rm MBE}$ itself as it is computationally
infeasible to calculate $E_{\rm MBE}$ due to the fact that it is
computationally infeasible to calculate the required von Neumann
entropies of reduced density operators given the
infinite-dimensional bases of the harmonic oscillators comprising
$|\psi_{\rm CM} \rangle$. This is so as these are generally
calculated by first diagonalizing $\rho$ and it is computationally
infeasible to do this, in general, when $\rho$ is a square matrix
of infinite dimension.

%
%calculate S_{L}
%

We begin with the initial density operator
$\rho_{\rm CM} = | \psi_{\rm CM}\rangle \langle \psi_{\rm CM}|$ which can be written
in the centre-of-mass number-state basis as
\begin{equation}
\rho_{\rm CM}=\sum_{{\cal N},{\cal N}'}^{\infty,\infty} f({\cal N},{\cal N}')
| {\cal N} \rangle_{1}|{\cal N} \rangle_{2} \;_{2}\langle {\cal N}'| _{1}\langle{\cal N}' |,
\end{equation}
where $f({\cal N},{\cal N}') = \tanh^{{\cal N} + {\cal N}'} r /
\cosh^{2} r$. To obtain a general $\rho_{Q_{j}}$, we trace over
the first $T$ atoms in the first FORT and the first $V$ in the
second one, arriving at
\begin{equation} \label{trace_real_calc}
\rho_{Q_{j}} = \sum_{\vec{P}=\vec{0}}^{\vec{\infty}}
\sum_{{\cal N},{\cal N}'}^{\infty,\infty}
f({\cal N},{\cal N}') \langle \vec{P} | {\cal N} \rangle_{1} |{\cal N} \rangle_{2}\;
_{2}\langle {\cal N}'| _{1}\langle {\cal N}' |\vec{P} \rangle,
\end{equation}
where $\vec{P}$ is a dummy variable given by $\vec{P} =
(p_{1}^{(1)},p_{2}^{(1)}...p_{T}^{(1)}, p_{1}^{(2)}... p_{V}^{(2)}
)$, where $p_{\alpha}^{(j)}$ denotes a vibrational number state
for the $\alpha^{\rm th}$ atom in the $x$ direction in the $j^{\rm
th}$ FORT, $\vec{0} = (0_{(1)},0_{(2)},0_{(3)}...0_{(T+V)})$ and
$\vec{\infty}=
(\infty_{(1)},\infty_{(2)},\infty_{(3)}...\infty_{(T+V)})$, where
a bracketed subscript enumerates the elements of $\vec{0}$ or
$\vec{\infty}$. We adopt a notation such that a sum of the form
$\sum_{\vec{X}=\vec{0}}^{\vec{Y}}$, where $\vec{X}$  and $\vec{Y}$
are the $F$-component vectors $(X_{1},X_{2} \ldots X_{F})$ and
$(Y_{1},Y_{2} \ldots Y_{F})$ respectively, denotes the set of sums
$\sum_{X_{1}=0}^{Y_{1}} \sum_{X_{2}=0}^{Y_{2}}
\sum_{X_{3}=0}^{Y_{3}} \ldots \sum_{X_{F}=0}^{Y_{F}}$.
Furthermore, we also assume that a state of the form $|\vec{X}
\rangle$ denotes the state $| X_{1} \rangle \otimes | X_{2}
\rangle \ldots | X_{F} \rangle$. Note that due to an exchange
symmetry for atoms in the same group of atoms, it is sufficient to
just consider the reduced density operators denoted by
Eqn~(\ref{trace_real_calc}) to deal with all possible
$\rho_{Q_{j}}$'s. That is, we do not need to consider, say,
tracing over the first and third atoms in the first FORT and the
second one in the second FORT. This is so as the $\rho_{Q_{j}}$
this yields is identical to that produced by tracing over the
first two atoms in the first FORT and the first one in the second
FORT.

%
%find S_{L}
%

We now find $[\rho_{Q_{j}}]^{2}$ and then trace over the remaining $2N - (T+V)$ atoms, producing
\begin{eqnarray} \label{reduce_trace} \nonumber
Tr([\rho_{Q_{j}}]^{2}) & = &
Tr \left( \sum_{\vec{P}=\vec{0}}^{\vec{\infty}}
\sum_{\vec{{\cal P}}=\vec{0}}^{\vec{\infty}}
\sum_{{\cal N},{\cal N}',{\cal M},{\cal
M}'}^{\infty,\infty,\infty,\infty} \right. \\ \nonumber
& &  \;\;\;\;\;\left. f({\cal N},{\cal N}') f({\cal M},{\cal M}') \right. \\ \nonumber
& & \times \left. \langle \vec{P} | {\cal N} \rangle |{\cal N} \rangle
\langle {\cal N}' | \langle {\cal N}' | \vec{P} \rangle \right. \\
& & \times \left. \langle \vec{{\cal P}} | {\cal M} \rangle | {\cal M} \rangle
\langle {\cal M}' | \langle {\cal M}' |\vec{{\cal P}} \rangle \right),
\end{eqnarray}
where, in analogy with $\vec{P}$, $\vec{{\cal P}}$ is a dummy
variable given by $\vec{{\cal P}} =  ({\cal P}_{1}^{(1)},{\cal
P}_{2}^{(1)}...{\cal P}_{T}^{(1)}, {\cal P}_{1}^{(2)}... {\cal
P}_{V}^{(2)} )$ where ${\cal P}^{(j)}_{\alpha}$ denotes a
vibrational number state in the $x$ direction for the $\alpha^{\rm
th}$ atom in the $j^{\rm th}$ FORT.

%
%calculate S_{L}
%

\begin{figure}[ht]
\center{\epsfig{figure=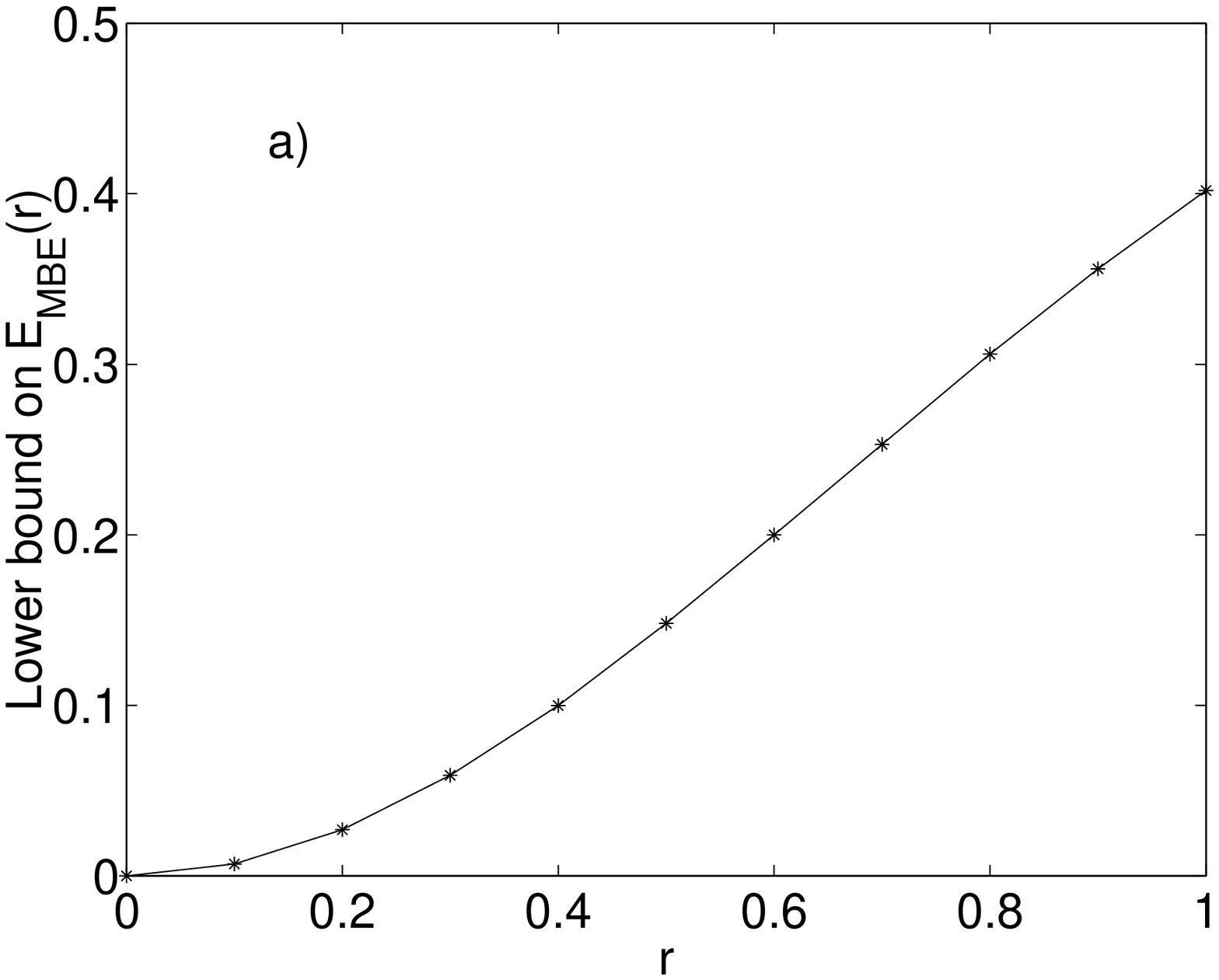,width=80mm}}
\center{\epsfig{figure=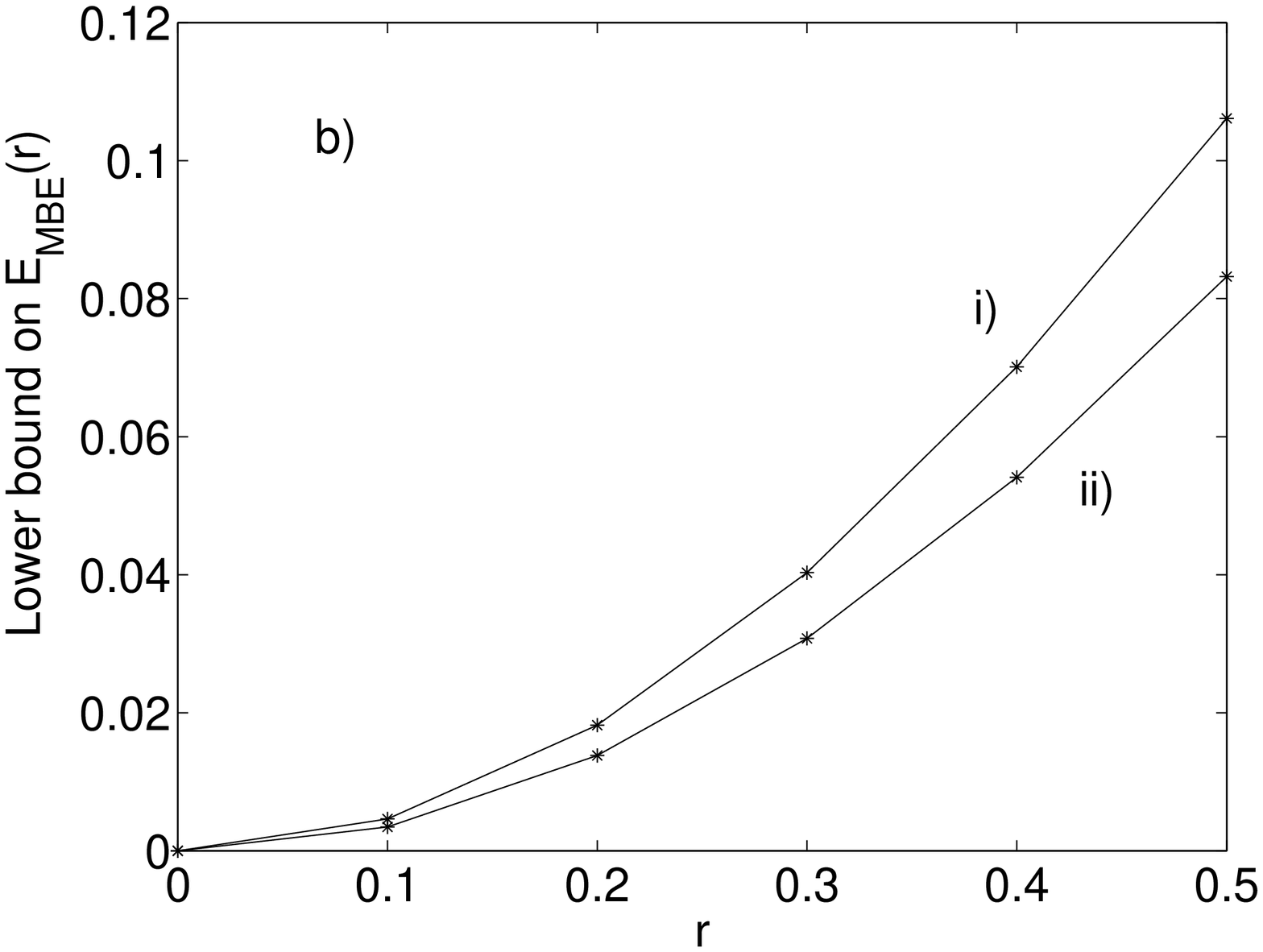,width=80mm}}
\caption{\label{ch5:fig1} Plots of lower bounds on $E_{\rm MBE}$
(dimensionless) for $| \psi_{\rm CM} \rangle$ as a function of $r$
(dimensionless) for a) $N=2$ (4-way entanglement), b) (i) $N=3$
(6-way entanglement) and (ii) $N=4$ (8-way entanglement). Note
that in all figures we have linearly interpolated between points
0.1 units apart on horizontal axes. Numerical errors are less than
$10^{-3}$ for all data points.}
\end{figure}

Using Eqn~(\ref{reduce_trace}), we now numerically determine
$S_{L}(\rho_{Q_{j}})$ for arbitrary $T$ and $V$ particular values of $N$ and $r$.
Our results provide lower bounds for $S(\rho_{Q_{j}})$ as
$S_{L}(\rho)/  \log_{2} {\rm e} \leq S(\rho)$
as can be verified by considering a power series expansion for $S(\rho)$.
Hence, knowing $S_{L}(\rho_{Q_{j}})$ for all bipartite splits of $| \psi_{\rm CM} \rangle$
allows us to infer a lower bound for ${\rm min}({\bf S_{\rm all}})$
and hence one for $E_{\rm MBE}$.
We thus calculate all $S_{L}(\rho_{Q_{j}})$ for $N=2,3,4$ for a range of
$r$ values numerically
using straightforward C++ code. These results are then used to place lower bounds on
$E_{\rm MBE}(|\psi_{\rm CM}\rangle)$ for 4-way, 6-way and 8-way entanglement
which appear in Figs \ref{ch5:fig1}(a) and (b).

%
%errors
%

As $|\psi_{\rm CM} \rangle$ is the sum
of an infinite number of statevectors, to calculate $S_{L}$ in practice,
we truncate the sum over ${\cal N}$ in the definition of
$|\psi_{\rm CM} \rangle$ at a finite value.
This induces errors in
our lower bounds for $E_{\rm MBE}(| \psi_{\rm CM} \rangle)$
for which upper bounds can be derived.
For all data points in Figs~\ref{ch5:fig1} (a) and (b), the
errors on our lower bounds for
$E_{\rm MBE}(| \psi_{\rm CM} \rangle)$
have been calculated to be less than $10^{-3}$ and hence are negligible.

%
%interesting features of graphs
%

Two interesting features of Figs~\ref{ch5:fig1} (a) and (b) are that, firstly, for a given $r$
value our lower bound on $E_{\rm MBE}$ decreases for increasing $N$.
It is {\em possible} that we can understand this behaviour by observing that
for constant $r$ we initially have a fixed entanglement resource, namely the entangled output of the NOPA.
It is conceivable that the decrease under consideration results from
this fixed resource being spread amongst a larger number of subsystems
as we increase $N$ thus, perhaps, causing it to distribute less bipartite
entanglement to any given bipartite split of $|\psi_{\rm CM}\rangle$.
In turn, this may decrease the $S_{L}$ of both halves of an arbitrary split, thus
explaining the decrease in our lower bound for $E_{\rm MBE}$ for increasing $N$.
The second interesting feature of Figs~\ref{ch5:fig1} (a) and (b) is that as we increase $r$,
$E_{\rm MBE}$ increases as expected given that an increased $r$ means
that we have more centre-of-mass entanglement.

\section{Discussion} \label{discuss}

%
%general
%

Throughout the paper, we have emphasised that $|\psi_{\rm CM}
\rangle$ contains $2N$-way entanglement. However, for this
entanglement to be meaningful it must have observable effects. One
feature of the system under consideration that makes its
entanglement conducive to producing such effect is the fact that
the atoms in the system are spatially separated and thus, in
principle, individually accessible. Thus, for example, we could
shine a sufficiently narrow laser beam on one of the atoms and,
provided it did not propagate perpendicular to the $x$-axis,
implement a local displacement on the vibrational state of the
atom in the $x$ direction. Furthermore, accessing individual atoms
is made easier by the fact that neighbouring atoms do not have be
located at successive cavity-field nodes. Instead, they can occupy
every second, third etc. node, thus increasing their spatial
separation and making it easier to address them one at a time.
Another advantageous consequence of the fact that each atom is
individually accessible is that it permits us to perform
measurements on the vibrational states of single atoms, perhaps by
employing a quantum-optical technique used to measure the position
of individual trapped atoms by having them interact strongly with
a low-photon number cavity mode \cite{hood00}.

%
%what we can do with \psi_{CM}
%

In light of the considerations of the previous paragraph, some
possible applications of the entanglement in $|\psi_{\rm CM}
\rangle$ are as follows:
\begin{enumerate}
\item {\em Violations of inequalities based on local realism.}
A number of such inequalities for an arbitrary number of quantum
systems have been formulated \cite{multi_party_bell}. Given the
close connection between violations of these inequalities and
entanglement, $|\psi_{\rm CM} \rangle$ is the sort of state we
might expect to violate at least some $2N$-party inequalities
based on local realism. However, as the Hilbert space for the
vibrational motion each atom is infinite-dimensional and not
two-dimensional as is the case for qubits, the violations may
require discretising or ``binning" measurement results of a
continuous variable such as quadrature phase amplitude.
%Explicitly verifying this, however, would
%be involved as calculating the quantum averages present in the
%inequalities would, most probably, be complicated.
\item {\em Solving quantum communication complexity problems (distributed quantum
computing).} Quantum communication complexity problems
\cite{grover97} involve a number of parties attempting to evaluate
some function $f$ for a particular input string. Each party is
given part of the input string and then uses shared prior
entanglement, local classical computation and public communication
in attempting to evaluate $f$. In such a scenario, the prior
entanglement can allow the evaluation to be performed in a
superior manner to that attainable classically. As the
entanglement in $|\psi\rangle_{\rm CM}$ is such that every atom in
the corresponding system is with every other one, it is a quantum
resource seemingly well-suited to being of use to in solving
quantum communication complexity problems better than can be done
classically.
\item{\em Continuous-variable quantum computation.}
Continuous-variable quantum computation \cite{lloyd99} involves
quantum computing with infinite-dimensional quantum systems as
opposed to the usual two-dimensional qubits. The most obvious way
to perform this sort of computation with the system under
consideration would be to, firstly, consider each atom in it as a
qudit in the limit of $d \rightarrow \infty$. After this, we would
then need to implement two-qudit gates by having different atoms
interact with each other in a pairwise manner. One method by which
this might be accomplished is by using a scheme \cite{deutsch00}
employed in optical lattices to get two spatially-separated
trapped neutral atoms of different species to interact with one
another. This is done by varying the polarisations of the
electromagnetic fields trapping the atoms which has the effect of
varying the potentials that the atoms see in such a manner that
they move towards each other. Once together, the atoms interact
via a dipole-dipole coupling. It is conceivable that this method
could be applied to implement two-qudit gates in the system of
interest. One complication, however, in utilising this scheme is
that it necessitates that we modify our system by having the $2N$
atoms in it comprised of two different species, perhaps with the
species of atom alternating as we move along each linear
configuration. Nevertheless, whilst the system under consideration
may not be the most natural one in which to do continuous-variable
quantum computation, there is some possibility that the
entanglement in it could be used to do this.
\end{enumerate}

%
%qualitative
%

\subsection{Qualitative results}
The thinking underlying {\bf Definition 1} is the same as that which underlies
the NPT sufficient condition for $M$-way entanglement.
However, there are significant differences between the two.
Firstly, {\bf Definition 1} involves arbitrary dimensional subsystems,
whereas the NPT sufficient condition
deals only with qubits. Secondly, the NPT sufficient condition
is a sufficient but not a necessary condition for $M$-way entanglement whereas the satisfaction of
{\bf Definition 1} is both necessary and sufficient for pure states.
Thirdly, the NPT sufficient condition uses
the partial transpose to determine the presence of $M$-way entanglement, whereas {\bf Definition 1} uses the
mathematically simpler entity the trace of the square of a reduced density operator.
Observe that {\bf Definition 1} is narrower than the
NPT sufficient condition in the sense that it only applies to
pure states whilst the NPT sufficient condition is
applicable to both pure and mixed states.

\subsection{Quantitative results}

%
%issues E_{2N}
%

A number of issues surround $E_{\rm MBE}$, which we now discuss.
\begin{enumerate}
\item {\em What does $E_{\rm MBE}$ tell us about what
quantum resource we have?} Ideally, we would like to be able to
relate $E_{\rm MBE}$ to one or more quantum tasks or protocols
such as distributed quantum computation with $E_{\rm MBE}$ telling
us something valuable about how well we can perform these tasks.
This is so as if we could do this, then it would increase $E_{\rm
MBE}$'s utility. Unfortunately, however, this has not yet been
accomplished.
\item {\em Can we tractably calculate $E_{\rm MBE}$?}
For an entanglement measure to be useful, it
must be tractable and able to be calculated in practice.
Unfortunately, $E_{\rm MBE}$ seems to be difficult to
calculate, at least for the state considered.
\end{enumerate}
Although $E_{\rm MBE}$ has the two above negative features we note
that, firstly, further research may eliminate them and, secondly,
we should consider them alongside the positive features of $E_{\rm
MBE}$ which are that it is a reasonable measure and that it helps
us to understand the nature of the entanglement in $| \psi_{\rm
CM} \rangle$ and also the capabilities of quantum state exchange.
Our results contribute to the understanding of multipartite
entanglement involving massive particles and infinite-dimensional
Hilbert spaces within a context that is not overly experimentally
infeasible.

%
%conclusion
%

To conclude, we have shown that quantum state exchange can be used
to produce the state $| \psi_{\rm CM} \rangle$ for two sets of
trapped atoms in spatially separated FORTs. We have also show that
$| \psi_{\rm CM} \rangle$ is a $2N$-way entangled state and, in
addition, have placed a lower bound on the amount of such
entanglement that it possesses. Finally, we have discussed quantum
information processing tasks that the $2N$-way entanglement in $|
\psi_{\rm CM} \rangle$ could be used to help perform.

\section{Acknowledgements}
DTP would like to thank Drs Scott Parkins, Bill Munro, Tobias
Osborne and Tim Ralph for valuable discussions. He would also like
to thank the referee for highlighting a shortcoming in the orginal
version and, finally, Dr Derrick Siu.

\end{document}